\documentclass[aps,prl,superscriptaddress,twocolumn,showpacs]{revtex4}
\usepackage{amsmath}
\usepackage{amssymb}
\usepackage{textcomp}
\usepackage{graphicx}
\usepackage{color}

\usepackage{soul,xcolor}
\setstcolor{red}

\input{epsf}

\begin{document}

\title{Coulomb-promoted spintromechanics in magnetic shuttle devices}

\author{O. A. Ilinskaya}
\email{ilinskaya@ilt.kharkov.ua} \affiliation{B. Verkin Institute
for Low Temperature Physics and Engineering of the National
Academy of Sciences of Ukraine, 47 Nauki Ave., Kharkiv 61103,
Ukraine}
\author{D. Radic}
\affiliation{Department of Physics, Faculty of Science, University
of Zagreb, Bijenicka 32, Zagreb 10000, Croatia}
\author{H. C. Park}
\email{hcpark@ibs.re.kr} \affiliation{Center for Theoretical
Physics of Complex Systems, Institute for Basic Science (IBS),
Daejeon 34051, Republic of Korea}
\author{I. V. Krive}
\affiliation{B. Verkin Institute for Low Temperature Physics and
Engineering of the National Academy of Sciences of Ukraine, 47
Nauki Ave., Kharkiv 61103, Ukraine}  \affiliation{Physical
Department, V.N. Karazin National University, Kharkiv 61077,
Ukraine}
\author{R. I. Shekhter}
\affiliation{Department of Physics, University of Gothenburg,
SE-412 96 G{\" o}teborg, Sweden}
\author{M. Jonson}
\affiliation{Department of Physics, University of Gothenburg,
SE-412 96 G{\" o}teborg, Sweden}

\date{\today}

\begin{abstract}
Exchange forces on the movable dot ("shuttle") in a magnetic
shuttle device depend on the parity of the number of shuttling
electrons. The performance of such a device can therefore be tuned
by changing the strength $U$ of Coulomb correlations to block or
unblock parity  fluctuations. We show that by increasing $U$ the
spintro-mechanics of the device crosses over, at $U=U_c(T)$, from
a mechanically stable regime to a regime of spin-induced shuttle
instabilities. This is due to enhanced spin-dependent mechanical
forces as parity fluctuations are reduced by a Coulomb blockade of
tunneling and demonstrates that single-electron manipulation of
single-spin controlled nano-mechanics is possible.
\end{abstract}
%\pacs{}

\maketitle
\noindent

Single-electronics \cite{devoret} and spintronics \cite{zutic} are
mesoscopic research areas related to two fundamental
properties of electrons: their charge and their spin. Strong
Coulomb correlations and quantum coherent electron spin
dynamics in nanometer-size conductors make them
candidates for future device applications. In this context it is
interesting to explore the interplay between spin- and
charge degrees of freedom on the nanometer length scale.

Tunneling injection of electrons into a nanoconductor is an obvious
way to control the amount of both charge and spin accumulated in
a nanometer scale spatial domain. However, in contrast to the amount of electric charge
the amount of electron spin that can be accumulated by this process is limited.
This is because while electrons with different spin projections can be
injected into the conductor, the net spin accumulated depends ---
assuming a spin-degenerate electronic spectrum ---
on the parity of the number of injected electrons.
The net accumulated spin,  at equilibrium, is at most equal to a single
electron spin and this occurs only for an odd number of
injected electrons.
Quantum fluctuations of the electron number destroy all effects
originating from parity, thus prohibiting the tunneling
accumulation of a finite average amount of spin. By suppressing these parity fluctuations
the Coulomb blockade phenomenon enhances the probability for
a finite spin to be accumulated.
This opens an intriguing
possibility to use the interplay between single-electronic and
spintronic properties for designing the functionality of nanoconductors.

Spintromechanics \cite{pulkin} relies on a coupling between
mechanical degrees of freedom and the electron spin in magnetic
nanoelectromechanical (NEM) devices \cite{zant, mceuen} (see e.g.
reviews Ref.~\onlinecite{blencowe}, 
Ref.~\onlinecite{aldridge}). The coupling is due to the magnetic
exchange interaction between spins accumulated in the
movable part of the NEM device (a metal grain or molecule here
called a ``dot") and the magnetization in the leads. This makes
spintromechanical phenomena an important tool for probing the spin
accumulated in a nanoconductor. One can therefore expect a
prominent role for Coulomb correlations in the spintromechanical
performance of magnetic NEM devices.

Below we consider the interplay between spintromechanical and
single-electron performances of a magnetic NEM system, taking the
magnetic shuttle device ( see, e.g., Refs.~\onlinecite{erbe},
\onlinecite{pistolesi}) as an example. We demonstrate that a dramatic
change of the mechanical behavior of the shuttle device can be
induced by using a gate to increase the electron-number (parity)
fluctuations in the dot, corresponding to a lifting of the Coulomb
blockade of tunneling. As a consequence of the related increase of
the fluctuations of the spin-dependent mechanical force on the dot
the shuttle instability of the magnetic NEM device, predicted to
occur in the absence of parity fluctuations in
Ref.~\onlinecite{kulinich}, is suppressed.

\begin{figure}
\begin{center}
\includegraphics[width=0.9\linewidth]{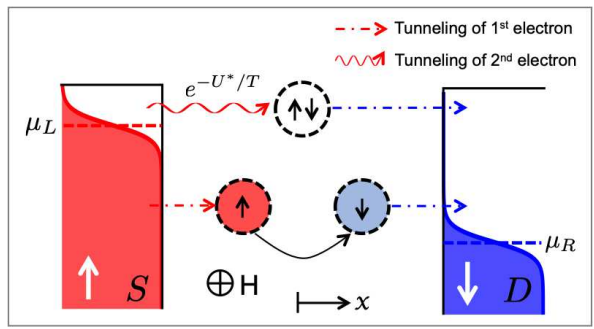}
\caption{
 Mechanism of Coulomb-promoted spintromechanics: A spin-up electron
loaded to the shuttle from the fully polarized source electrode
($S$) interacts with the magnetizations in the leads to create a
magnetic exchange force that attracts the dot to the source (red
dot). An external magnetic field ``rotates" the spin into a
spin-down state, thereby reversing the sign of the exchange force
so that the shuttle is pushed towards the drain ($D$) (blue dot).
In the case of a full Coulomb blockade ($U^\ast \rightarrow
\infty$, see text), which prevents double occupation, the shuttle
becomes mechanically active and energy is pumped into the
mechanical shuttle vibrations. With a partial Coulomb blockade
double occupation of the shuttle occurs with probability
$\exp(-U^\ast/T)$, where $T$ is the temperature,
leading to electron parity fluctuations, which are detrimental to
the energy pumping mechanism. } \label{fig1}
\end{center}
\end{figure}

A typical magnetic shuttle device comprises two magnetic metallic
electrodes, which form a standard tunnel junction, while a movable
small conductor (dot) is trapped
(e.g., by van der Waals forces) in the tunneling region of the
device. By biasing the device by a voltage difference or a temperature
gradient, a flow of electrons is induced between the magnetic electrodes with
the possibility for extra electrons to be resident  in the
dot. In a steady state, the electric charge and net electron spin
accumulated in the dot interact with the electric field (caused by
the bias voltage) and with the magnetic (exchange) forces (caused
by the interaction of the quantum dot spin with the magnetizations
of the leads). This is how a coupling between mechanical
vibrations and electron tunneling through the device is induced (see supplementary material).

%In the analysis below we neglect the voltage-dependent electric force on 
%the dot in comparison with the magnetic exchange force. This is a good approximation if 
%the electric field ${\cal E}$ acting on the shuttle is sufficiently weak.If ${\cal E}$ is 
%increased there is at some point a transition from the regime of spintromechanical  shuttling
%discussed here to electromechanical shuttling for which the parity effect is obviously irrelevant.
%A description of this transition is beyond the scope of the present paper.

We model the quantum dot by a single spin-degenerate electron level and assume that the leads 
are fully spin-polarized half-metals with anti-parallel magnetizations.
In Fig.~1 different tunneling and spin-flip events, which modify
the electronic population in the dot, are specified. Tunneling
events, leading to a singly and a doubly occupied dot, are
discriminated by the extra Coulomb energy cost $U$
for the dot to be doubly occupied. If the
energy of the singly occupied dot is $\varepsilon_0$, then the
activation energy $U^\ast$ required for a second electron to
tunnel to the dot is
\begin{equation}
\label{U*} U^\ast = U + \varepsilon_0-\mu_L,
\end{equation}
where $\mu_L$ is the chemical potential in the left
lead.

Let us first consider the limit of strong Coulomb correlations ($U
\to \infty$), in which case double occupation of the dot is
prohibited. Then, we note that an important requirement for the
proper operation of a magnetic shuttle device is that an external
magnetic field is applied perpendicular to the magnetizations of
the leads. This field makes it possible for a spin-up electron
injected into the dot from the source electrode, to flip its spin,
as indicated in Fig.~1. The resulting change of spin orientation
enables the exchange force to push the dot away from the source
towards the drain electrode, into which the extra electron
tunnels, thus causing the empty dot to move back to the source
electrode. This is the mechanism by which an electron flow through
the device can generate mechanical oscillations and, under certain
conditions, the spin-induced shuttle instability predicted in
Ref.~\onlinecite{kulinich} (where it was  assumed that no more
than one electron occupies the dot at any time).

Next we explore the consequences of our ability to vary the effect
of the Coulomb correlations by increasing the temperature or by
gating the device. To that end we consider strong Coulomb blockade
regime, $U^\ast \gg T$, where $T$ is the temperature, in which case there is a small probability
 for a spin-up second electron to tunnel into the dot already occupied
 by a spin-down electron. Three electronic configurations on the dot are
 now possible. Two of them, the dot singly occupied by a spin-up electron --- prevented to tunnel
 into the drain since it has only spin-down states --- and the dot doubly occupied
  by one spin-up and one spin-down electron, are ``mechanically inactive"
 in the sense that no net work is done during one oscillation period
 against the exchange
 force. Only the third configuration, the dot singly occupied by a spin-down electron,
 is mechanically active, which
 enables energy, accumulated by electrons, to be
transferred into mechanical vibration energy.

The functionality of the magnetic shuttle device is determined by
the coupling of three different degrees of freedom. Those are
related to: (i) the spatial tunneling motion of electrons between
the leads via the quantum dot, (ii) the rigid mechanical motion of
the movable dot, which affects electron tunneling probabilities,
and (iii) the electron spin dynamics, which influences the
mechanical motion of the dot through the exchange force,
acting on the quantum dot.

Referring to Fig.~1, the
classical nanomechanics of the shuttle vibrations  can be
described by Newton's equation for the oscillator with a spin- and
displacement-dependent ``external" exchange force
\cite{ilinskaya},
\begin{equation}\label{x}
m\ddot x+m\omega^2
x=-\frac{\alpha}{2}\left[\rho_\uparrow\{x(t)\}-\rho_\downarrow\{x(t)\}\right].
\end{equation}
Here the coefficient $\alpha$ is the magnitude of the exchange
force per unit spin experienced by a shuttle, situated in the
middle of the gap ($x$=0) between the oppositely magnetized leads,
where the exchange energy $J(x)=J_L(x)-J_R(x)\simeq -\alpha
x$ [we consider a magnetically symmetric contact,
$J_L(0)=J_R(0)\equiv J$]; $J$ is the strength of exchange
interaction, $m$ is the mass of the dot and $\omega$ is the
angular frequency of the mechanical vibrations of the dot.

The exchange force on the right-hand side of Eq.~(\ref{x}) is
proportional to the displacement-dependent amount of spin
accumulated in the dot, which depends on the difference between
the probabilities $\rho_{\uparrow(\downarrow)}\{x(t)\}$ for the
dot to be singly occupied by a spin-up (down) electron. These
probabilities are solutions to a complex kinetic problem for the
quantum evolution of the electron density operator $\hat\rho$,
describing the interplay between mechanical vibrations,
coherent spin dynamics in the exchange and external magnetic
fields and incoherent tunneling of electrons \cite{SM}.

The corresponding equations can be derived to lowest order in the
tunneling probabilities by following the procedure used in Ref.
\onlinecite{ilinskaya}. In this approach electrons in the leads
are described by equilibrium distribution functions and therefore
all electronic degrees of freedom in the leads are easily averaged
out. The electron distribution in the dot is described using the
Fock representation. In this representation, four eigenstates,
corresponding to the empty dot, $|0\rangle$, to the
dot, singly occupied by a spin-up (down) electron,
$\mid\uparrow\rangle$ ($\mid\downarrow\rangle$), and to the doubly
occupied dot, $|2\rangle$, form a
complete Hilbert space for the single-level quantum dot. The
matrix elements of the density operator form a $6$-vector with
components: $\rho_0\equiv\langle 0|\hat\rho_d|0\rangle$, $\rho_\uparrow\equiv\langle\uparrow|\hat\rho_d|\uparrow\rangle$, $\rho_\downarrow\equiv\langle\downarrow|\hat\rho_d|\downarrow\rangle$, $\rho_{\uparrow\downarrow}=\rho_{\downarrow\uparrow}^\ast\equiv\langle\uparrow|\hat\rho_d|
\downarrow\rangle$, $\rho_2\equiv\langle 2|\hat\rho_d|2\rangle$.
Here $\rho_0$ ($\rho_2$) is the probability for the dot to be
empty (doubly occupied),
while the other matrix elements correspond to a singly occupied
dot (including the non-diagonal components
$\rho_{\uparrow\downarrow}$, $\rho_{\downarrow\uparrow}$). The
number of independent variables can be reduced by one using the
normalization condition
$\rho_0+\rho_\uparrow+\rho_\downarrow+\rho_2=1$. All matrix elements 
experience two types of dynamical evolution:
(i) electron tunneling events, described by classical ``collision"
integrals, and (ii) quantum coherent spin evolution in response to
the exchange field and to the external magnetic field $H$. In
order to study the mechanical motion of the quantum dot, we are
interested in the dynamics of the ''spin active" linear
combination of distribution functions,
$\rho_\uparrow-\rho_\downarrow$.
 It is easy to show that the symmetric
spin-neutral quantities $R_0=\rho_0+\rho_2$ and
$\rho_\uparrow+\rho_\downarrow=1-R_0$ are decoupled from the
equations for the four other linear combinations: $R_1=\rho_0-\rho_2$, $R_2=\rho_\uparrow-\rho_\downarrow$, $R_3=-i(\rho_{\uparrow\downarrow}-\rho_{\uparrow\downarrow}^\ast)$, and $R_4=\rho_{\uparrow\downarrow}+\rho_{\uparrow\downarrow}^\ast$.
%\begin{eqnarray}\label{4-vector}
%R_1&=&\rho_0-\rho_2,\quad
%\hspace{0.95cm} R_2=\rho_\uparrow-\rho_\downarrow,\nonumber\\
%R_3&=&-i(\rho_{\uparrow\downarrow}-\rho_{\uparrow\downarrow}^\ast),\quad
%R_4=\rho_{\uparrow\downarrow}+\rho_{\uparrow\downarrow}^\ast.
%\end{eqnarray}
It follows that the set of equations, describing tunneling and spin evolution
of the density operator, can be written in the compact form \cite{SM}

\begin{equation}\label{EqR}
\frac{d\overrightarrow{R}}{dt}=\left(\hat A_\Gamma+\hat A_H+\hat
A_J\right)\overrightarrow{R}+\overrightarrow{B}.
\end{equation}
Here the matrices $\hat A_{\Gamma,H,J}$, which
describe the dynamics, caused by tunneling ($\hat A_\Gamma$), spin
evolution in the external magnetic field ($\hat A_H$), and spin
evolution due to the exchange
interaction ($\hat A_J$),
and the ``source term" $\overrightarrow{B}$ are defined by
Eqs.~(25)-(31) of Ref.~\onlinecite{SM}.

Equations (\ref{x}) and (\ref{EqR}) form a closed set of
equations, describing the spintromechanics of the magnetic shuttle
device. Being nonlinear, they can in a general case only be solved
numerically. Whether a nanomechanical instability can be triggered
by injecting an electron current into the device is the most
important question to address. The answer can be obtained by
linearizing Eqs.~(\ref{x}) and (\ref{EqR}) with respect to small
mechanical dot displacements and by finding the condition for the
exponential growth (in time) of their amplitude. Having in mind
the above qualitative analysis, one expects that the criterion for
a mechanical instability crucially depends on the strength of the
Coulomb charging energy $U$. In the Coulomb blockade regime, $U^\ast\gg
 T$, the existence of a shuttle instability for
such a system was predicted in Ref.~\onlinecite{ilinskaya}. The
solution of the linearized Eqs.~(\ref{x}) and (\ref{EqR}) can also
be found in the opposite limit of non-interacting electrons,
$U=0$, for the case of a symmetric junction
$\Gamma_L=\Gamma_R\equiv\Gamma$ \cite{SM} ($\Gamma_{L,R}$ are
partial dot tunneling widths). In this case we have shown
[\onlinecite{SM}] that the exchange interaction, caused by the
electron spin, results in a positive imaginary part of the
renormalized angular frequency $\Omega$ of the mechanical
vibrations. This corresponds to an exponential time decay of the
amplitude of mechanical vibrations [$x(t)\sim\exp(i\Omega t)] $].
Straightforward calculations \cite{SM} yield for the
rate of change, $r$, of the amplitude of the nanomechanical vibrations
 at $U=0$
\begin{eqnarray}\label{Im}
r=-\text{Im}\Omega(h)&=&-\frac{\alpha^2}{16m}\sum_{j=L,R}
\frac{f_j(\varepsilon_0-h)-f_j(\varepsilon_0+h)}{h}\times\nonumber\\
&\times&\frac{\Gamma(\Gamma^2+\omega^2+4h^2)}{(\Gamma^2-\omega^2+4h^2)^2+4\omega^2\Gamma^2}.
\end{eqnarray}
Here $\hbar=1$, $h=g\mu_B H/2$ is the external magnetic field in
energy units ($\mu_B$ is the Bohr magneton, $g$ is the
gyromagnetic ratio). The negative value of the rate
 is evident from the monotonic decay of
the Fermi distribution function, $f_j(\epsilon)=1/\left\{\exp[(\epsilon-\mu_j)/T_j+1]\right\}$,
%\begin{equation}\label{FermiFunctions}
%f_j(\epsilon)=\left\{1+\exp[(\epsilon-\mu_j)/T_j]\right\}^{-1}
%\end{equation}
determined by chemical potentials $\mu_j$ and temperatures $T_j$
in the leads $j=L,R$. In the numerical
analysis below we assume that the temperatures in the leads are
the same, $T_L=T_R=T$, and in the unbiased device $\mu_L=\mu_R=\varepsilon_F$.

The result, presented in
Eq.~(\ref{Im}), suggests that a crucial change in
spintromechanical performance occurs when Coulomb correlations are
tuned. While increasing $U$ from zero, one stimulates a
performance, which eventually results in the occurrence of a
nanomechanical instability at a certain critical value of
$U=U_c$.

The temperature-dependent crossover from a mechanically stable
regime, where spontaneous dot fluctuations are damped out, to a
shuttling regime, where they are amplified, occurs at
$U^\ast_c$ 
when the ratio of double to single spin-down
occupations reaches a critical value
$\rho_2/\rho_{\downarrow}=n_c$. The critical value $U^\ast_c$ and
$n_c$ are related by a simple steady state relation
$\Gamma_L\exp(-U^\ast/T)=n_c\Gamma_R$ (here we neglect electron
backflow and we omit all terms proportional to small magnetic field). 
The critical activation energy reads
\begin{equation}\label{U_c}
 U_c^\ast\simeq T\ln\left(\frac{\Gamma_L}{n_c\Gamma_R}\right),
\end{equation}
corresponding to a linear dependence of the critical
Coulomb energy on temperature and a slight (logarithmic) decrease of the slope with
an increase of the ``asymmetry parameter`` $\gamma=\Gamma_R/\Gamma_L$ of the shuttle device.

%
%%%%%%%%%%%%%%%%%%%%%%%%%%
\begin{figure}
\begin{center}
\includegraphics[width=\columnwidth]{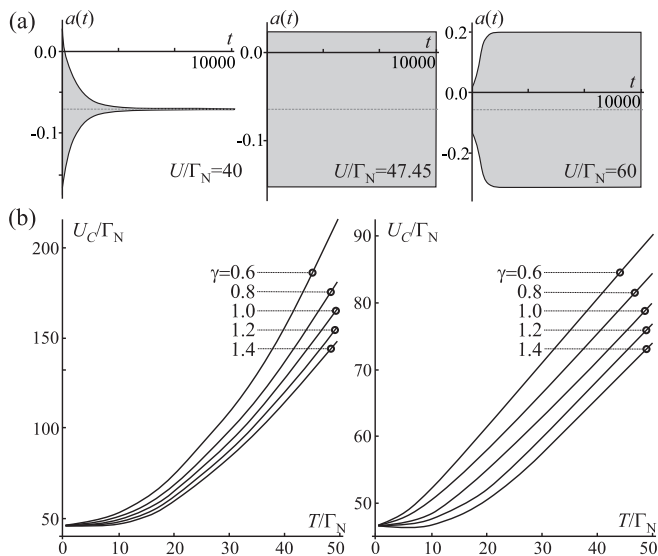}
\caption{ Characteristics of the mechanical dot vibrations as
determined by Eqs. (\ref{x}) and (\ref{EqR}) for the NEM device
sketched in Fig.~1: (a) Amplitude $a(t)$ of the dot vibrations (in
units of the tunneling length $\lambda$) as a function of time $t$
(in units of $\hbar/\Gamma_N$, where
$\Gamma_N=\Gamma_L\Gamma_R/(\Gamma_L+\Gamma_R)$, $\Gamma_{L,R}$ are the widths of
the energy level on the dot) for three different values of the
Coulomb blockade energy $U$ at $T/{\Gamma_N}=5$. For small values
of $U$ spontaneous vibrations are damped out (left panel), while
for large $U$ they are amplified and develop into sustained
finite-amplitude vibrations (right panel). A crossover between the
two regimes occurs at $U=U_c$, when the vibration amplitude stays
constant over time (middle panel). (b) Temperature dependence of
the critical value of the Coulomb blockade energy $U_c(T)$ for a symmetric
voltage biased device $\mu_L-\varepsilon_F = \varepsilon_F-\mu_R$
(left panel) and a device with a strongly biased drain electrode
$\mu_L-\varepsilon_F\sim T, \varepsilon_F-\mu_R\gg T$ (right
panel) for different values of the asymmetry parameter
$\gamma=\Gamma_R/\Gamma_L$. The results were obtained for the
following parameters: left lead bias $(\mu_L-\varepsilon_F) =
50\,\Gamma_N$, vibration frequency $\omega = 1\,\Gamma_N/\hbar$,
external magnetic field $H = 0.5\,\Gamma_N/\mu_B$, magnetic
exchange energy $J=1.5\,\Gamma_N$, dot level detuning energy
$(\varepsilon_0-\varepsilon_F)=2\,\Gamma_N$, and the
electromechanical coupling constant $\kappa=\hbar^2J/(m
l\lambda\Gamma_N^2)=0.09$, where $\kappa$ is the coefficient
multiplying r.h.s of Eq. (\ref{x}) 
%$\left[\rho_\uparrow\{x(t)\}-\rho_\downarrow\{x(t)\}\right]$ if
written in terms of dimensionless variables $x$
and $t$. For $\Gamma_N =$\,0.5\,meV and $\lambda = l =$\,0.1\,nm
one finds that $m \sim 2\cdot 10^{-25}$\,kg and $\omega \sim 2\pi
\times 0.1$\,THz are in reasonable agreement with experiments on
single C$_{60}$ molecules trapped between metal electrodes
\cite{mceuen,park}. } \label{Numerics-pict}
\end{center}
\end{figure}
%%%%%%%%%%%%%%%%%%%%%%%%%%%%
%

In order to analyze shuttle vibrations in the steady state,
eventually reached after a shuttle instability, linearizing the
problem with respect to dot displacements is not an adequate
approach. Instead, one has to deal with the full nonlinear and
nonlocal in time spintromechanical problem at hand numerically. In
Fig.~\ref{Numerics-pict} we present results of numerical solutions
to the coupled equations for the time-development of the density
matrix components and of the mechanical oscillations of the dot,
which are valid for arbitrary (not only small) dot displacements.
In this case Eq.~(\ref{x}) has to be generalized by replacing the
spintromechanical exchange force $\alpha= (2J/l)$ by a
coordinate-dependent force, $\alpha \rightarrow
(2J/l)\cosh\left(x(t)/l\right)$; $l$ is the decay length of the
exchange interaction, and we consider a magnetically symmetric
junction [\onlinecite{SM}].

Results of numerical simulations of the nonlinear and nonlocal temporal
dynamics of the mechanical vibrations
are presented in Fig.~2(a). One can readily see that, depending on
how the Coulomb correlation energy $U$ is related to a critical value $U_c$,
spontaneous small-amplitude vibrations are either damped out ($U < U_c$,
left panel), maintained at a constant amplitude ($U = U_c$,
middle panel) or amplified until they reach some steady-state amplitude $a_0$
($U > U_c$, right panel).

It is remarkable that in the event of a shuttle instability [such
as the one illustrated in the right panel of Fig.~2(a)] the
mechanical vibration amplitude saturates even though no
phenomenological friction term is included in Eq.~(\ref{x}). The physical
explanation of this ``self-saturation" effect is based on the fact
that the electron shuttling phenomenon relies on
retardation effects in the mechanical subsystem. These effects
disappear in the limit of oscillation amplitudes that are large
enough for the dot to come so close to the source and drain
electrodes that the tunneling rate of charge relaxation becomes
higher than the mechanical oscillation frequency.

In Fig.~2(b) (left panel) we present the numerically obtained temperature
dependence of $U_c$ for different asymmetry parameters
$\gamma$  and a symmetric biasing of the device,
$\mu_L-\varepsilon_F=\varepsilon_F-\mu_R$.
It is clearly seen from the figure that the linear temperature
dependence of $U_c$, which is expected from the qualitative analysis,
see Eqs.~(\ref{U_c}) and (\ref{U*}), holds rather accurately in the temperature range
$25 < T/\Gamma_N < 45$. At higher temperatures the deviation from
linearity becomes more pronounced due to a strong suppression of
electron shuttling caused by electron parity fluctuations.
An obvious reason for that is temperature stimulated population (ignored in
Eq.~(\ref{U_c})) of
electronic states in the drain (right) electrode.
This  impedes electron tunneling to the drain lead thus suppressing the
electric current through the device and diminishing the power supply to the mechanical
vibrations. Such a blocking effect can be removed in
asymmetrically biased device when $(\varepsilon_F-\mu_R)/T\gg 1$,
$(\mu_L-\varepsilon_F)/T\sim 1$.
The results of numerical simulations in
this case are presented in Fig.~2(b), right panel. One can clearly see an accurate linear dependence
in a large temperature interval in full agreement with Eq. (\ref{U_c}). Moreover,
using the value $n_c\simeq 0.28$ obtained in our numerical analysis,
one finds from Eqs.~(\ref{U_c}) and (\ref{U*}) that the slope $s(n_c,\gamma)\equiv dU_c/dT$
for $\gamma=1$ is $ s(\gamma=1)\simeq 1.27 $.
This  can be compared with the slope $s\simeq 1.0$
of the curve plotted for $\gamma=1$
in Fig.~2(b) (right panel). We conclude that there is good agreement between
the exact numerical result and what we anticipated from our qualitative picture
of Coulomb promoted magnetic-shuttle spintromechanics.

In conclusion, we have shown that Coulomb correlations play an
important role for the nanomechanical properties of a magnetic
shuttle device due to their ability to trigger a spintromechanical
shuttle instability.  Such an instability occurs when the Coulomb
blockade charging energy exceeds a critical value, which depends
on temperature and the strength of the enabling external magnetic
field. The effect opens a possibility for single-electronic
manipulation of spintromechanical performance.

\textbf{\emph{Acknowledgement}}. This work was supported by the
Institute for Basic Science in Korea (IBS-R024-D1); the Swedish
Research Council (VR) ;  the Croatian Science Foundation, project
IP-2016-06-2289, and by the QuantiXLie Centre of Excellence, a
project cofinanced by the Croatian Government and the European
Union through the European Regional Development Fund - the
Competitiveness and Cohesion Operational Programme (Grant
KK.01.1.1.01.0004). The authors acknowledge the hospitality of
PCS IBS in Daejeon (Korea).

\begin{widetext}

\centerline{\bf \Large{Supplemental Material}}

\vspace{10mm} %10mm vertical space

The Hamiltonian of our spintromechanical system (see Fig.~1) consists of 3 terms
\begin{equation}\label{H0}
\hat H=\hat H_l+\hat H_d+\hat H_t,
\end{equation}
where $\hat H_l$ describes noninteracting spin-polarized electrons in the leads
 ( we assume that the leads are
fully spin-polarized half-metals with anti-parallel
magnetizations)
\begin{equation}\label{Hl}
\hat H_l=\sum_{k,j}\varepsilon_{k,j}a_{k,j}^\dag
a_{k,j},\qquad j=L,R=\uparrow,\downarrow.
\end{equation}
Here $a_{k,j}^\dag$ $(a_{k,j})$ is the creation (annihilation)
operator for an electron with  
momentum $k$ in 
lead $j$. We model the quantum dot by a single
spin-degenerate electron level. The Hamiltonian of the
quantum dot (QD) reads
\begin{equation}\label{Hd}
\hat H_d=\sum_\sigma\varepsilon_\sigma c_\sigma^\dag c_\sigma
-h\left(c_\uparrow^\dag
c_\downarrow+c_\downarrow^\dag
c_\uparrow\right)+Uc_\uparrow^\dag c_\uparrow
c_\downarrow^\dag c_\downarrow+\hat H_m,
\end{equation}
\begin{equation}\label{Hm}
\hat H_m=\frac{p^2}{2m}+\frac{m\omega^2}{2}x^2,
\end{equation}
where $\varepsilon_\sigma=\varepsilon_0-(\sigma/2)J(x)$ is the
spin- ($\sigma=\uparrow,\downarrow=+,-$) and
position-dependent energy of the quantum dot spin-split levels
($\varepsilon_0$ is the level energy, $J(x)=J_L(x)-J_R(x)$ is the
coordinate-dependent magnetic exchange energy per unit QD spin between the QD and the
leads); $c_\sigma^\dag$ $(c_\sigma)$ is the creation
(annihilation) operator for electron with spin projection
$\sigma=\uparrow,\downarrow$ in the dot; $h\equiv g\mu_B H/2$ and
$H$ is the external magnetic field, which is perpendicular to the
plane of magnetization in the leads ($g$ is the gyromagnetic ratio,
$\mu_B$ is the Bohr magneton), $U$ is the electron-electron
repulsion energy. Vibrations of the dot are described by the
Hamiltonian $H_m$ of a harmonic oscillator. In what follows we will
treat $x$ and $p$ as classical variables.

In Eq.(\ref{H0})-(\ref{Hd}) and in the analysis below we neglect the voltage-dependent electric
force on the dot in comparison with the magnetic exchange force.
This is a good approximation if the electric field ${\cal E}$
acting on the
 shuttle is sufficiently weak.
If ${\cal E}$ is increased there is at some point a transition from the
regime of spintromechanical  shuttling discussed
here to electromechanical shuttling for which the parity effect is obviously irrelevant.
A description of this transition is beyond the scope of the present paper.

The tunneling of electrons between lead $j=L,R=-,+$ and a movable
QD is described by a tunneling Hamiltonian with coordinate-dependent
tunneling amplitude $t_j(x)=t_j\exp(jx/\lambda)$, $\lambda$
is the characteristic tunneling length, 
\begin{equation}\label{Ht}
\hat H_t=t_L(x)\sum_k c^\dag_\uparrow a_{k,L}+t_R(x)\sum_k
c^\dag_\downarrow a_{k,R}+\text{h.c.}
\end{equation}

We solve the problem of mechanical instability in our NEM system
by using the density operator method. The density operator obeys the von Neumann
equation $(\hbar=1)$
\begin{equation}\label{vN}
i\partial_t\hat\rho(t)=[\hat H, \hat\rho(t)].
\end{equation}

In what follows we consider the regime of sequential electron
tunneling in the NEM transistor ($\text{max}\{T,
\mu_L-\mu_R\}\gg\Gamma$, where $\Gamma$ is the width of the electron
energy level in the dot, $T$ is the temperature, $\mu_L$, $\mu_R$
are the chemical potentials in the leads). Then the total density
operator is factorized,
$\hat\rho(t)\approx\hat\rho_d(t)\otimes\hat\rho_l$, where
$\hat\rho_d(t)$ is the QD density operator and $\hat\rho_l$ is the
equilibrium density matrix of the leads. By tracing out the %leads
leads'
degrees of freedom it is straightforward to derive %the 
a set of
equations for the matrix elements of $\hat\rho_d$ in the Fock
space of a single level QD: $\rho_0=\langle
0|\hat\rho_d|0\rangle$,
$\rho_\uparrow=\langle\uparrow|\hat\rho_d|\uparrow\rangle$,
$\rho_\downarrow= \langle\downarrow|\hat\rho_d|\downarrow\rangle$,
$\rho_{\uparrow\downarrow}=\langle\uparrow|\hat\rho_d|\downarrow\rangle=
\rho_{\downarrow\uparrow}^\ast$, $\rho_2=\langle
2|\hat\rho_d|2\rangle$. One finds that
\begin{equation}\label{EqRho}
|\dot\rho_j\rangle=\hat A_\rho|\rho_j\rangle,
\end{equation}
%
%Here 
where the $6\times 6$ matrix $\hat A_\rho$ has the %following 
form
\begin{equation}\label{ARho}
\hat A_\rho=
\left(\begin{array}{cccccc}
A_{11} &
A_{12} &
A_{13} &
A_{14} &
A_{14}^\ast &
0
\\
A_{21} &
A_{22} &
0 &
A_{24} &
A_{24}^\ast &
A_{26}
\\
A_{31} &
0 &
A_{33} &
A_{34} &
A_{34}^\ast &
A_{36}
\\
A_{41} &
A_{42} &
A_{43} &
A_{44} &
0 &
A_{46}
\\
A_{41}^\ast &
A_{42}^\ast &
A_{43}^\ast &
0 &
A_{44}^\ast &
A_{46}^\ast
\\
0 &
A_{62} &
A_{63} &
A_{64} &
A_{64}^\ast &
A_{66}
\\
\end{array}\right),
\end{equation}
%
%where 
with the following matrix elements $A_{ij}$:
\begin{eqnarray}\label{A11}
A_{11}&=&-\Gamma_L(x)f_L^{+}-\Gamma_R(x)f_R^{+}-\Upsilon_{1L}(x)+\Upsilon_{1R}(x),\; A_{12}=\Gamma_L(x)(1-f_L^{+})-\Upsilon_{1L}(x),\\
A_{13}&=&\Gamma_R(x)(1-f_R^{+})+\Upsilon_{1R}(x),\;A_{14}=-\Upsilon_{2L}(x)-\Upsilon_{2R}(x),\\
A_{21}&=&\Gamma_L(x)f_L^{+}+\Upsilon_{1L}(x),\;A_{22}=-\Gamma_L(x)(1-f_L^{+})-\Gamma_R(x)f_R^{U,+}+\Upsilon_{1L}(x)+\Upsilon_{1R}^{U}(x),\\
A_{24}&=&-i h+\Upsilon_{2L}(x)+\Upsilon_{2R}^{U}(x),\;A_{26}=\Gamma_R(x)(1-f_R^{U,+})+\Upsilon_{1R}^{U}(x),\\
A_{31}&=&\Gamma_R(x)f_R^{+}-\Upsilon_{1R}(x),\;A_{33}=-\Gamma_R(x)(1-f_R^{+})-\Gamma_L(x)f_L^{U,+}-\Upsilon_{1L}^{U}(x)-\Upsilon_{1R}(x),\\
A_{34}&=&i h+\Upsilon_{2R}(x)+\Upsilon_{2L}^{U}(x),\;A_{36}=\Gamma_L(x)(1-f_L^{U,+})-\Upsilon_{1L}^{U}(x),\\
A_{41}&=&\Upsilon_{2L}(x)+\Upsilon_{2R}(x),\; A_{42}=-i h+\Upsilon_{2L}(x)+\Upsilon_{2R}^{U}(x),\; A_{43}=i h+\Upsilon_{2R}(x)+\Upsilon_{2L}^{U}(x),\\
A_{44}&=&i J(x)-\frac{1}{2}\Gamma_L(x)(1-f_L^{+}+f_L^{U,+})-\frac{1}{2}\Gamma_R(x)(1-f_R^{+}+f_R^{U,+})\nonumber\\
&-&\frac{1}{2}\left[\Upsilon_{1L}(x)-\Upsilon_{1L}^{U}(x)-\Upsilon_{1R}(x)+\Upsilon_{1R}^{U}(x)\right],\\
A_{46}&=&\Upsilon_{2L}^{U}(x)+\Upsilon_{2R}^{U}(x),\; A_{62}=\Gamma_R(x)f_R^{U,+}-\Upsilon_{1R}^{U}(x),\;A_{63}=\Gamma_L(x)f_L^{U,+}+\Upsilon_{1L}^{U}(x),\\
A_{64}&=&-\left[\Upsilon_{2L}^{U}(x)+\Upsilon_{2R}^{U}(x)\right],\;A_{66}=-\Gamma_L(x)(1-f_L^{U,+})-\Gamma_R(x)(1-f_R^{U,+})+\Upsilon_{1L}^{U}(x)-\Upsilon_{1R}^{U}(x).
\label{A66}
\end{eqnarray}
In Eqs.~(\ref{A11})--(\ref{A66}) we introduced the following
notations,
\begin{eqnarray}\label{Ups-1}
\Upsilon_{1L/R}(x)&=&f_{L/R}^{-}\frac{J(x)\Gamma_{L/R}(x)}{\sqrt{J^2(x)+4h^2}},\;\Upsilon_{1L/R}^{U}(x)=f_{L/R}^{U,-}\frac{J(x)\Gamma_{L/R}(x)}{\sqrt{J^2(x)+4h^2}},\\
\Upsilon_{2L/R}(x)&=&f_{L/R}^{-}\frac{h\Gamma_{L/R}(x)}{\sqrt{J^2(x)+4h^2}},\;\Upsilon_{2L/R}^{U}(x)=f_{L/R}^{U,-}\frac{h\Gamma_{L/R}(x)}{\sqrt{J^2(x)+4h^2}};
\label{Ups-2}
\end{eqnarray}
here
\begin{eqnarray}\label{f-pm}
2f_{L,R}^{\pm}&=&f_{L,R}(E_{-})\pm f_{L,R}(E_{+}),\;
2f_{L,R}^{U,\pm}=f_{L,R}(E_{-}+U)\pm f_{L,R}(E_{+}+U),
\label{f-Upm}
\end{eqnarray}
and
\begin{equation}\label{Epm}
E_{\pm}=\varepsilon_0\pm\frac{\sqrt{J^2(x)+4h^2}}{2},
\end{equation}
\begin{equation}\label{FermiFunctions}
f_j(\epsilon)=\left\{1+\exp[(\epsilon-\mu_j)/T_j]\right\}^{-1}.
\end{equation}
We use the system of equations (\ref{EqRho}) for numerical
calculations (see main text).

For analytical calculations it is convenient to use linear
combinations of $\rho_j$. Then the symmetric spin-neutral
combination $R_0=\rho_0+\rho_2$ (and
$\rho_\uparrow+\rho_\downarrow=1-\rho_0-\rho_2$) is decoupled from
%four other 
the other four quantities $R_1=\rho_0-\rho_2$,
$R_2=\rho_\uparrow-\rho_\downarrow$,
$R_3=-i(\rho_{\uparrow\downarrow}-\rho_{\uparrow\downarrow}^\ast)$,
$R_4=\rho_{\uparrow\downarrow}+\rho_{\uparrow\downarrow}^\ast$.
The system of equations for $R_j$ $(j=1-4)$ takes the form
\begin{equation}\label{EqR}
|\dot R(t)\rangle=\hat A\{x(t),U\}|R(t)\rangle+|B\{x(t),U\}\rangle,
\end{equation}
with
$\hat A\{x(t),U\}=\hat A_\Gamma\{x(t),U\}+\hat A_H\{x(t),U\}+\hat A_J\{x(t),U\}$.
Here we introduced matrices related to tunneling (subindex $\Gamma$), the external
magnetic field (subindex $H$) and the exchange interaction (subindex $J$)
\begin{equation}\label{AGamma}
\hat A_\Gamma\{x(t),U\}=-\frac{1}{2}
\left(\begin{array}{cccc}
 F_{+}^{+,-}(x,U) & -F_{-}^{-,+}(x,U) & 0 & 0 \\
 -F_{-}^{+,-}(x,U) & F_{+}^{-,+}(x,U) & 0 & 0 \\
 0 & 0 & F_{+}^{-,+}(x,U) & 0 \\
 0 & 0 & 0 & F_{+}^{-,+}(x,U) \\
\end{array}\right),
\end{equation}
\begin{equation}\label{Ah}
\hat A_H\{x(t),U\}=
\left(\begin{array}{cccc}
 0 & 0 & 0 & -H_{+}^{-}(x,U) \\
 0 & 0 & 2h & H_{-}^{-}(x,U) \\
 0 & -2h & 0 & 0 \\
 H_{+}^{-}(x,U) & H_{-}^{-}(x,U) & 0 & 0 \\
\end{array}\right),
\end{equation}
\begin{equation}\label{AJ}
\hat A_J\{x(t),U\}=\frac{1}{2}
\left(\begin{array}{cccc}
 -J_{-}^{-}(x,U) & -J_{+}^{-}(x,U) & 0 & 0 \\
 J_{+}^{-}(x,U) & J_{-}^{-}(x,U) & 0 & 0 \\
 0 & 0 & -J_{-}^{-}(x,U) & J(x) \\
 0 & 0 & -J(x) & J_{-}^{-}(x,U) \\
\end{array}\right),
\end{equation}
\begin{equation}\label{B}
|B\{x(t),U\}\rangle=\frac{1}{2}
\left(\begin{array}{c}
F_{+}^{-,-}(x,U)-J_{-}^{+}(x,U)\\
-F_{-}^{-,-}(x,U)+J_{+}^{+}(x,U)\\
0\\
2H_{+}^{+}(x,U)\\
\end{array}\right).
\end{equation}
Here we denote
\begin{equation}\label{F}
F_{\pm}^{j,k}(x,U)=\Gamma_{\pm}(x)+j\left[\Gamma_L(x)f_L^{+}\pm\Gamma_R(x)f_R^{+}\right]+
k\left[\Gamma_L(x)f_L^{U,+}\pm\Gamma_R(x)f_R^{U,+}\right],
\end{equation}
with $j,k=(+,-)$
\begin{equation}\label{H}
H_{\pm}^{\eta}(x,U)=\left\{\Gamma_L(x)f_L^{-}\pm\Gamma_R(x)f_R^{-}+
\eta\left[\Gamma_L(x)f_L^{U,-}\pm\Gamma_R(x)f_R^{U,-}\right]\right\}\frac{h}{\sqrt{4h^2+J^2(x)}},
\end{equation}
\begin{equation}\label{J}
J_{\pm}^{\eta}(x,U)=\left\{\Gamma_L(x)f_L^{-}\pm\Gamma_R(x)f_R^{-}+
\eta\left[\Gamma_L(x)f_L^{U,-}\pm\Gamma_R(x)f_R^{U,-}\right]\right\}\frac{J(x)}{\sqrt{4h^2+J^2(x)}};
\end{equation}
and $\eta=\pm 1$. Notice that
$H_{\pm}^{\eta}(x,U)/h=J_{\pm}^{\eta}(x,U)/J(x)$. In Eq.~(\ref{F}) the
notation
%\begin{equation}\label{Gamma-pm}
$\Gamma_{\pm}(x)=\Gamma_L(x)\pm\Gamma_R(x)$
%\end{equation}
is introduced.

In the case of non-interacting electrons, $U=0$, the analytic
solution can be simplified for a symmetric tunnel junction,
$\Gamma_L(x=0)=\Gamma_R(x=0)=\Gamma$, $J_L(x=0)=J_R(x=0)$. We
solve the system (\ref{EqR}) by %perturbations,
perturbation theory with $R_i(t)\approx
R_i^{0}+R_i^{1}(t)$, $R_i^{1}(t)\propto x(t)$, assuming the
displacement $x$ to be small. Then the equations for $R_1^{1}$ and
$R_4^{1}$ are decoupled from the equations for $R_2^{1}$ and
$R_3^{1}$. Therefore the analysis of the mechanical instability in
this particular case is reduced to a simpler problem --- one has
to solve two coupled linear equations
\begin{equation}\label{Eq-U0}
\left(\begin{array}{c}
\dot R_2^{1}\\
\dot R_3^{1}
\end{array}\right)=
\left(\begin{array}{cc}
-\Gamma & 2h\\
-2h & -\Gamma
\end{array}\right)
\left(\begin{array}{c}
R_2^{1}\\
R_3^{1}
\end{array}\right)-\alpha(f_L^{-}+f_R^{-})
\left(\begin{array}{c}
\Gamma/(2h)\\
1
\end{array}\right)
x(t),
\end{equation}
where $\alpha>0$ (the exchange force per unit QD spin) is the derivative of the exchange
energy per unit QD spin, $J(x)\approx -\alpha x$ ({\it Cf.} Fig. 1). Substituting the solution
$R_2^{1}$ into Eq.~(2) of the main text, we obtain
the desired Eq.~(6).

\clearpage

\end{widetext}


\begin{thebibliography}{99}
\bibitem{devoret}
{\it Single Charge Tunneling} (H.~Grabert, M.H.~Devoret, Eds.),
Plenum, New York (1992).

\bibitem{zutic}
I.~Zutic, J.~Fabian, S.~Das Sarma, {\it Rev. Mod. Phys.} {\bf 76},
323 (2004).

\bibitem{pulkin}
R.I.~Shekhter, A.~Pulkin, M.~Jonson, {\it Phys. Rev. B} {\bf 86},
100404(R) (2012).

\bibitem{zant}
M.~Poot, H.S.J.~van der Zant, {\it Phys. Rep.} {\bf 511}, 273
(2012).

\bibitem{mceuen}
Abhay N. Pasupathy, Radoslaw C. Bialczak, Jan Martinek, Jacob E. Grose,
Luke A. K. Donev, Paul L. McEuen, and Daniel C. Ralph {\it Science} {\bf 306}, 86
(2004).

\bibitem{blencowe} M.P.~Blencowe, Contemp. Phys.
{\bf 46}, 249 (2005).

\bibitem{aldridge} J.S.~Aldridge, A.N.~Cleland,
R.~Knobel, D.R.~Schmidt, C.S.~Yung, Proceedings of Spie --- the
International Society for Optical Engineering {\bf 4591}, 11
(2001).

\bibitem{erbe} A.~Erbe, C.~Weiss, W.~Zwerger,
R.H.~Blick, Phys. Rev. Lett. {\bf 87}(9), 096106 (2001).

\bibitem{pistolesi} F.~Pistolesi, Phys. Rev.~B {\bf
69}(24), 245409 (2004).

\bibitem{kulinich}
S.I.~Kulinich, L.Y.~Gorelik, A.N.~Kalinenko, I.V.~Krive,
R.I.~Shekhter, Y.W.~Park, and M.~Jonson, {\it Phys. Rev. Lett.}
{\bf 112}, 117206 (2014).

\bibitem{ilinskaya}
O.A.~Ilinskaya, S.I.~Kulinich, I.V.~Krive, R.I.~Shekhter,
H.C.~Park, and M.~Jonson, {\it New J. Phys.} {\bf 20}, 063036
(2018).

\bibitem{SM}
See Supplemental Material at [URL will be inserted by publisher] for details.

\bibitem{park}
Hongkun Park, Jiwoong Park, Andrew K. L. Lim, Erik H. Anderson, A. Paul Alivisatos, and Paul L. McEuen,
Nature {\bf 407}, 58 (2000).

\newpage
\clearpage


\end{thebibliography}
\end{document}